\documentclass[final,5p,times,twocolumn]{elsarticle}

\usepackage[T1]{fontenc}
\usepackage[utf8]{inputenc}

\usepackage{amsmath}
\usepackage{amsfonts}
\usepackage{amssymb}
\usepackage{amsxtra}
\usepackage{array}
\usepackage{color}
\usepackage{dcolumn}
\usepackage{graphicx}
\usepackage{hepunits}
\usepackage{xspace}
\usepackage{CJKutf8}

\definecolor{purple}{rgb}{0.5,0,0.5}
\definecolor{blue}{rgb}{0.0,0,0.9}
\definecolor{prdblue}{rgb}{0.133,0.118,0.498}
\usepackage[colorlinks=true, pdfstartview=FitV, linkcolor=prdblue, citecolor= prdblue, urlcolor=prdblue]{hyperref}

\usepackage[mathscr,scaled=1.15]{urwchancal}
\DeclareFontFamily{OT1}{pzc}{}
\DeclareFontShape{OT1}{pzc}{m}{it}%
{<-> s * [1.15] pzcmi7t}{}
\DeclareMathAlphabet{\mathpzc}{OT1}{pzc}{m}{it}

\biboptions{sort&compress}

\journal{Physics Letters B}

\newcommand{\beq}{\begin{equation}}
\newcommand{\eeq}{\end{equation}}
\newcommand{\ba}{\begin{array}}
\newcommand{\ea}{\end{array}}
\newcommand{\bea}{\begin{align}}
\newcommand{\eea}{\end{align}}
\newcommand{\bi}{\begin{itemize}}
\newcommand{\ei}{\end{itemize}}
\newcommand{\ben}{\begin{enumerate}}
\newcommand{\een}{\end{enumerate}}
\newcommand{\bc}{\begin{center}}
\newcommand{\ec}{\end{center}}
\newcommand{\bl}{\begin{flushleft}}
\newcommand{\el}{\end{flushleft}}
\newcommand{\br}{\begin{flushright}}
\newcommand{\er}{\end{flushright}}

\begin{document}
\begin{CJK*}{UTF8}{gbsn}


\begin{frontmatter}

\title{$\,$\\[-7ex]\hspace*{\fill}{\normalsize{\sf\emph{Preprint no}. NJU-INP 093/24}}\\[1ex]
%
Sketching pion and proton mass distributions}

\author[NKU]{Xiaobin Wang (王晓斌)%
$^ {\href{https://orcid.org/0000-0002-2786-5296}{\textcolor[rgb]{0.00,1.00,0.00}{\sf ID}},}$}

\author[NKU]{Zanbin Xing (幸瓒彬)%
$^ {\href{https://orcid.org/0000-0001-8910-6941}{\textcolor[rgb]{0.00,1.00,0.00}{\sf ID}},}$}

\author[NKU]{Lei Chang (常雷)%
$^ {\href{https://orcid.org/0000-0002-4339-2943}{\textcolor[rgb]{0.00,1.00,0.00}{\sf ID}},}$}

\author[NJU,INP]{\\Minghui Ding (丁明慧)%
    $^{\href{https://orcid.org/0000-0002-3690-1690}{\textcolor[rgb]{0.00,1.00,0.00}{\sf ID}},}$}

\author[UHe]{Kh\'epani Raya%
       $^{\href{https://orcid.org/0000-0001-8225-5821}{\textcolor[rgb]{0.00,1.00,0.00}{\sf ID}},}$}

\author[NJU,INP]{Craig D. Roberts%
       $^{\href{https://orcid.org/0000-0002-2937-1361}{\textcolor[rgb]{0.00,1.00,0.00}{\sf ID}},}$}

\address[NKU]{
School of Physics, \href{https://ror.org/01y1kjr75}{Nankai University}, Tianjin 300071, China}

\address[UHe]{Department of Integrated Sciences and Center for Advanced Studies in Physics, Mathematics and Computation, \href{https://ror.org/03a1kt624}{University of Huelva}, E-21071 Huelva, Spain.}

\address[NJU]{
School of Physics, \href{https://ror.org/01rxvg760}{Nanjing University}, Nanjing, Jiangsu 210093, China}
\address[INP]{
Institute for Nonperturbative Physics, \href{https://ror.org/01rxvg760}{Nanjing University}, Nanjing, Jiangsu 210093, China\\[1ex]
%
\href{mailto:leichang@nankai.edu.cn}{leichang@nankai.edu.cn} (LC);
\href{mailto:cdroberts@nju.edu.cn}{cdroberts@nju.edu.cn} (CDR)
\\[1ex]
Date: 2024 October 16\\[-6ex]
}

\begin{abstract}
A light-front holographic model is used to illustrate an algebraic scheme for constructing a representation of a hadron's zero-skewness generalised parton distribution (GPD) from its valence-quark distribution function (DF) and electromagnetic form factor, $F_H$, without reference to deeply virtual Compton scattering data.
The hadron's mass distribution gravitational form factor, $A_H$, calculated from this GPD is harder than $F_H$; and, for each hadron, the associated mass-density profile is more compact than the analogous charge profile, with each pion near-core density being larger than that of its proton partner.  These features are independent of the scheme employed.
\end{abstract}


\end{frontmatter}
\end{CJK*}

\section{Introduction}
\label{s1}
Hadron electromagnetic form factors (EFFs) are defined via the in-hadron expectation value of the electromagnetic current in quantum chromodynamics (QCD), which only involves quark contributions because gluons do not carry electric charge.
Owing to their capacity for revealing structural details of hadron targets and the longtime availability of electron accelerators, such EFFs have been a focus of scrutiny by experiment and theory for many years -- see, \emph{e.g}., Refs.\,\cite{Perdrisat:2006hj, Eichmann:2016yit, Brodsky:2020vco}.
Notwithstanding this attention, numerous questions pertaining to both nonperturbative and perturbative aspects of QCD remain unanswered; so, this interest continues today, with modern and anticipated high-luminosity, high-energy facilities expected to deliver related data whose reach in momentum-squared transfer may extend to $Q^2 \lesssim 30 m_p^2$ \cite{Roberts:2021nhw, Arrington:2021biu, Chen:2020ijn, Wang:2022xad, Carman:2023zke}, where $m_p$ is the proton mass.

Analogous gravitational form factors (GFFs) are defined via the in-hadron expectation value of the QCD energy momentum tensor.
Terrestrial experiments involving graviton + hadron scattering are unrealisable.
However, following development of the generalised parton distribution (GPD) formalism \cite{Ji:1996ek, Radyushkin:1996nd, Mueller:1998fv}, it was recognised that empirical information on hadron GFFs might become accessible via data on deeply virtual Compton scattering (DVCS), \emph{i.e}., by adapting electron beams to a new purpose \cite{Diehl:2003ny, Guidal:2013rya, Mezrag:2023nkp}.
Here, too, precise data are expected from modern and anticipated facilities.
Thus, a theory effort is underway, which focuses on the prediction and physical interpretation of hadron GFFs.
There have been many model studies -- see, \emph{e.g}., Refs.\,\cite{Polyakov:2018zvc, Azizi:2019ytx, Neubelt:2019sou, Xing:2022mvk, Shi:2023oll, Li:2023izn, Nair:2024fit, Liu:2024vkj}; and predictions for pion and proton GFFs have recently been obtained using both continuum and lattice Schwinger function methods \cite{Xu:2023izo, Yao:2024ixu, Hackett:2023rif, Hackett:2023nkr}.  These studies can potentially reveal links between EFFs and GFFs, and tie both to phenomena underlying and connected with the emergence of mass in strong interactions \cite{Roberts:2021nhw, Binosi:2022djx, Ding:2022ows, Ferreira:2023fva, Raya:2024ejx}.

A key feature of GPDs is the connection they provide between a hadron's parton distribution functions (DFs) and their EFFs and GFFs.  This suggests that even before sufficient precise DVCS data become available, one might use existing knowledge of DFs and EFFs to make projections for a hadron's GFFs.  Pursuing this idea, Ref.\,\cite{Xu:2023bwv} used pion + nucleus Drell-Yan and electron + pion scattering data \cite{Conway:1989fs, Horn:2007ug, Huber:2008id} to develop ensembles of model-independent representations of the pion GPD and therefrom arrive at a data-driven prediction for the pion mass distribution GFF, denoted $A_\pi(Q^2)$ below.  A principal conclusion of Ref.\,\cite{Xu:2023bwv} is that $A_\pi(Q^2)$ is less compact (harder) than $F_\pi(Q^2)$, its EFF.  Moreover, in general, defined in the usual way from these form factors, the associated radii are constrained as follows:
\begin{equation}
\label{radiibound}
\tfrac{1}{2} \leq (r_A^\pi/r_F^\pi)^2 \leq 1\,.
\end{equation}
The lower bound is saturated by a pointlike two valence body system and the upper by an infinitely massive such system, for which all radii must smoothly approach zero from above and, therefore, their ratio approaches unity.
Kindred analyses may be found in Refs.\,\cite{Chavez:2021llq, DallOlio:2024vjv}.

Herein, our goal is to introduce an algebraic simplicity to such analyses and sketch an extension to the proton. We accomplish this using a light-front holographic model of hadrons (LFHM) \cite{Sufian:2016hwn, deTeramond:2018ecg, Chang:2020kjj}.  Some generality and QCD connections are lost, but these costs are offset by the clarity and insights gained.

\section{GPDs in a Light-Front Holographic Model}
\label{s2}
Adopting the LFHM described in Ref.\,\cite{Sufian:2016hwn, deTeramond:2018ecg, Chang:2020kjj}, hadron electromagnetic form factors (EFFs) are expressed via the Euler Beta function:
{\allowdisplaybreaks
\begin{align}
\label{eq:EFFint}
    F_{\tau}(Q^2) & = \frac{1}{N_{\tau}} B\left(\tau-1,\frac{1}{2} + \frac{Q^2}{4\lambda}\right)\nonumber\\
    &= \frac{1}{N_{\tau}}\int_0^1(1-y)^{\tau-2}y^{Q^2/4\lambda-1/2}dy\,,
\end{align}
where, for ground states, $\tau$ is the number of constituent quarks and/or antiquarks in the hadron;
$Q^2$ is the momentum-squared transfer in the scattering process;
$N_{\tau}=\Gamma(1/2)\Gamma(\tau-1)/\Gamma(\tau-1/2)$ ensures unit normalisation at $Q^2=0$;
and $\lambda$ is the model mass scale.
In this case, for $2 \leq \tau \in \mathbb Z$,
\begin{equation}
\label{eq:EFFpoles}
    F_\tau(Q^2) = \frac{1}{(1 + Q^2/M_0^2)(1 + Q^2/M_1^2) \cdots (1 + Q/M_{\tau-2}^2)}\,,
\end{equation}
with $M_n^2 = 4\lambda (n+1/2)$.  Assuming that the pion ($\tau=2$) EFF may be approximated as a $Q^2$ monopole characterised by the $\rho$-meson mass, $m_\rho=0.775\,$GeV \cite{ParticleDataGroup:2024cfk}, then $\surd \lambda = 0.548\,$GeV.  Amongst this model's merits are its simplicity and reproduction of large-$Q^2$ power-law scaling behaviour for pion and nucleon electromagnetic form factors.  The model does not incorporate the $\ln Q^2$ scaling violations that are characteristic of QCD \cite{Lepage:1980fj}.
}

\begin{figure}[t]
\vspace*{0ex}

\leftline{\hspace*{0.5em}{{\textsf{A}}}}
\vspace*{-2ex}
\centerline{\includegraphics[width=0.9\columnwidth]{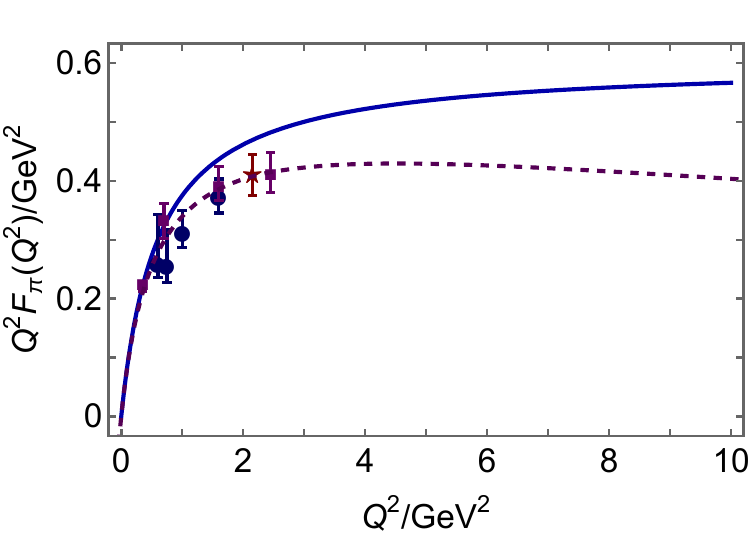}}
\vspace*{2ex}

\leftline{\hspace*{0.5em}{{\textsf{B}}}}
\vspace*{-2ex}
\centerline{\includegraphics[width=0.9\columnwidth]{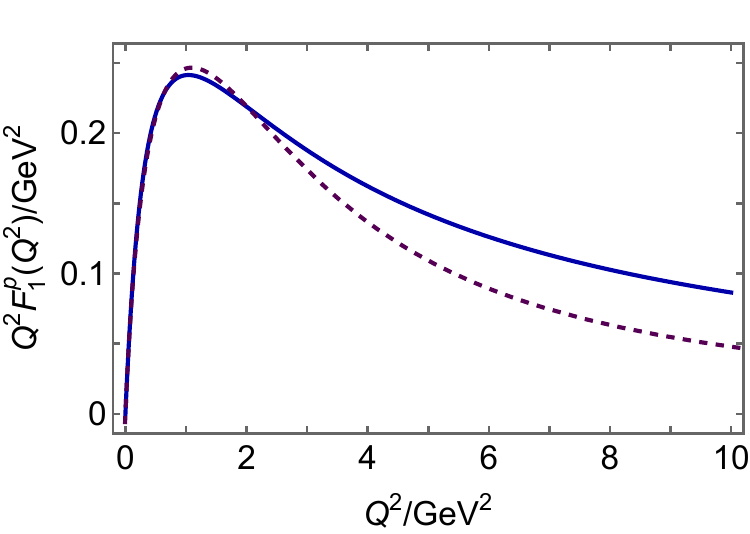}}

\caption{\label{EFFs}
{\sf Panel A}.
Pion form factor.
Solid blue curve -- LFHM, Eq.\,\eqref{eq:EFFpoles};
dashed purple curve -- prediction obtained using continuum Schwinger function methods (CSMs) \cite{Yao:2024drm};
data obtained and analysed by  JLab $F_\pi$ Collaborations \cite{Horn:2007ug, Huber:2008id}.
{\sf Panel B}.
Proton Dirac form factor.
Solid blue curve -- LFHM, Eq.\,\eqref{eq:EFFpoles};
and dashed purple curve -- CSM prediction \cite{Yao:2024uej}.  On the displayed domain, the CSM result is indistinguishable from a standard parametrisation of data \cite{Kelly:2004hm}.
}
\end{figure}

The quality of the LFHM's EFF descriptions is illustrated in Fig.\,\ref{EFFs}.
Regarding the pion, the omission of scaling violations entails that a good description of the form factor is only obtained on $Q^2 \lesssim 4 m_\rho^2$ -- see, \emph{e.g}., Ref.\,\cite{Yao:2024drm}.
Concerning the proton Dirac form factor, the picture is better because, in this case, scaling violations set in at larger values of $Q^2$.
The LFHM delivers a poor description of the neutron Dirac form factor \cite{Sufian:2016hwn}.
It is against this backdrop that one may judge the efficacy of the LFHM as a starting point for hadron physics phenomenology.

Turning now to QCD, the contribution of a given valence quark to a hadron's EFF may be expressed in terms of a GPD \cite{Diehl:2003ny}:
\begin{equation}
\label{eq:zerothmoment}
F_\tau^q(Q^2)= \int_0^1dx\, H_{{\rm v},\tau}^q(x,\xi, -Q^2)\,.
\end{equation}
Here we have limited ourselves to those terms which link to valence-quark distribution functions (DFs) in the forward limit ($\xi=0$, $Q^2=0$):
\begin{equation}
\label{DFGPD}
{\mathpzc q}_\tau(x) = H_{{\rm v},\tau}^q(x,0,0)\,.
\end{equation}
There is a similarity between Eqs.\,\eqref{eq:EFFint}, \eqref{eq:zerothmoment}, which we shall proceed to exploit.

Consider the following two functions, in general distinct:
\begin{equation}
\label{integralreps}
F(x;Q^2)  = \int_0^x \! dy\, f(y;Q^2) \,, \;
G(x;Q^2)  = \int_0^x \! dy\, g(y;Q^2) \,,
\end{equation}
where $f$, $g$ are integrable, non-negative and continuously differentiable ($C^1$) functions on $x\in (0,1)$.  Plainly, $F(0,Q^2)=0=G(0,Q^2)$.
The similarity between Eqs.\,\eqref{eq:EFFint}, \eqref{eq:zerothmoment} is recovered by requiring $F(1,Q^2)=G(1,Q^2)$.

Under the conditions stipulated, $F(x;Q^2)$ and $G(x;Q^2)$ are monotonically increasing $C^1$ functions of $x$, which are equal at the endpoints of their support domains.
(For this discussion, $Q^2$ is merely a label.)
It follows that there is another $C^1$ function $w(x;Q^2)$, satisfying
\begin{equation}
\label{wcons}
w(0;Q^2)=0\,, \quad w(1;Q^2)=1 \,, \quad w^\prime(x;Q^2)\geq0\,,
\end{equation}
where the derivative is on $x$, for which $F(w(x;Q^2);Q^2) =G(x;Q^2)$.
This is simply the statement that two functions, possessing the properties highlighted above, can be deformed, one into the other, by a $C^1$ coordinate mapping.
(The functions $F(x,\epsilon)=\ln(1+ \epsilon^2 x^2)/\ln(1+\epsilon^2)$, $\epsilon \in \mathbb R$, and $G(x)= \sin(\pi x/2)$ define a class of useful, illustrative examples.)
It follows that
\begin{equation}
f(w(x);Q^2) w^\prime (x;Q^2) = g(x;Q^2)\,.
\end{equation}

Returning to Eqs.\,\,\eqref{eq:EFFint}, \eqref{eq:zerothmoment} and using Eq.\,\eqref{DFGPD}, one may immediately infer that the LFHM GPD has the form
\begin{subequations}
\label{LFHMGPD}
\begin{align}
H_{{\rm v},\tau}^q(x,& 0, -Q^2) = {\mathpzc q}_\tau(x;Q^2) [w_\tau(x;Q^2)]^{Q^2/4\lambda} \,,
\label{dewA}\\
{\mathpzc q}_\tau(x;Q^2) & = \frac{1}{N_\tau}
[1-w_\tau(x;Q^2)]^{\tau-2}
 [w_\tau(x;Q^2)]^{-1/2} w_\tau^\prime (x;Q^2)\,. \label{dew}
\end{align}
\end{subequations}
This analysis means that when using the LFHM, the zero-skewness GPD is known once $w_\tau(x;Q^2)$ is determined.


A GPD with a product form kindred to that in Eq.\,\eqref{dewA} may be obtained from a factorised representation of a hadron's light-front wave function -- see the discussion in Ref.\,\cite{Raya:2021zrz}.  Even though factorised functions are poor pointwise approximations on $(x\simeq 0,1; \lambda/Q^2\simeq 0)$, they can nevertheless deliver reliable results for integrated quantities \cite{Xu:2018eii}.

Following Ref.\,\cite{Chang:2020kjj}, we proceed by supposing that $w_\tau (x;Q^2)$ is independent of $Q^2$, working hereafter with $w_\tau (x)$.
In this case, the profile function for a given hadron is completely determined by its valence quark DFs.
Such DFs depend on the resolving scale, $\zeta$.
Since the LFHM works with constituent-like degrees of freedom, then the natural choice is the hadron scale, $\zeta_{\cal H}$, whereat valence degrees of freedom carry all hadron properties.
This identification of $\zeta_{\cal H}$ is a basic tenet of the all-orders (AO) DF evolution scheme detailed in Ref.\,\cite{Yin:2023dbw}.
The AO approach has been employed successfully in numerous applications, \emph{e.g}., delivering unified predictions for all pion, kaon, and proton (unpolarised and polarised) DFs \cite{Cui:2020tdf, Lu:2022cjx, Cheng:2023kmt, Yu:2024qsd} and pion fragmentation functions \cite{Xing:2023pms}, and delivering a viable species separation of nucleon gravitational form factors \cite{Yao:2024ixu}.

 \begin{figure}[t]
    \centering
    \includegraphics[width=0.9\linewidth]{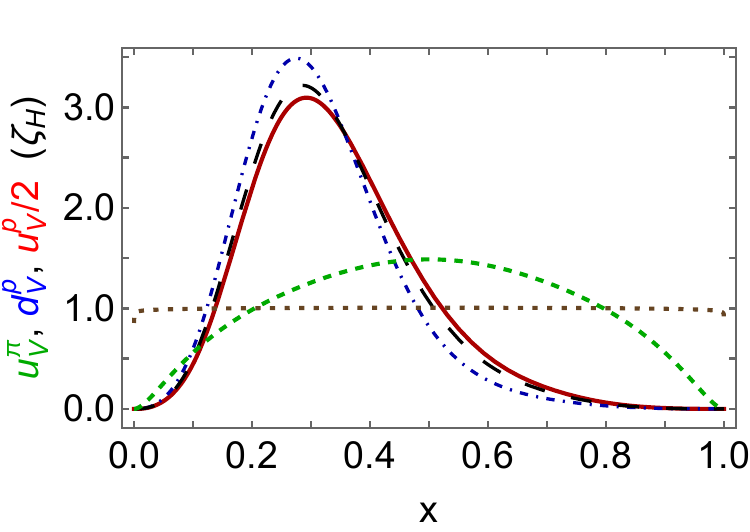}
    \caption{\label{DFinput}
    Hadron scale valence quark DFs for the pion \cite{Cui:2020tdf} (dashed green) and proton \cite{Lu:2022cjx} ($0.5 \times u$ quark -- solid red; $d$ -- dot-dashed blue).
    The long-dashed grey curve is
    ${\mathpzc q}_3(x;\zeta_{\cal H}) = [
    {\mathpzc u}_{\rm v}^p(x;\zeta_{\cal H})+{\mathpzc d}_{\rm v}^p(x;\zeta_{\cal H})]/3$.
    The DF in Eq.\,\eqref{SCIDF} is the dotted brown curve.
    }
\end{figure}

Hadron-scale valence-quark DFs for the pion and proton are available in Refs.\,\cite{Cui:2020tdf, Lu:2022cjx}.  Drawn in Fig.\,\ref{DFinput}, they were computed using continuum Schwinger function methods (CSMs) and are consistent with much available empirically-based information.  Notably, owing to correlations in the proton wave function and consistent with experiment \cite{Abrams:2021xum, Cui:2021gzg}, ${\mathpzc u}(x;\zeta_{\cal H}) \neq 2 {\mathpzc d}(x;\zeta_{\cal H})$.  This feature is not preserved in the LFHM unless additional model degrees-of-freedom and parameters are introduced \cite{deTeramond:2018ecg}.  To avoid that in determining $w_{\tau=3} (x)$, we use instead a mean valence quark DF in the proton, \emph{i.e}., ${\mathpzc q}_3(x;\zeta_{\cal H}) $ in Fig.\,\ref{DFinput}.  As will be shown below, this \emph{Ansatz} has no material impact on our discussion.

\begin{figure}[t]
    \centering
    \includegraphics[width=0.9\linewidth]{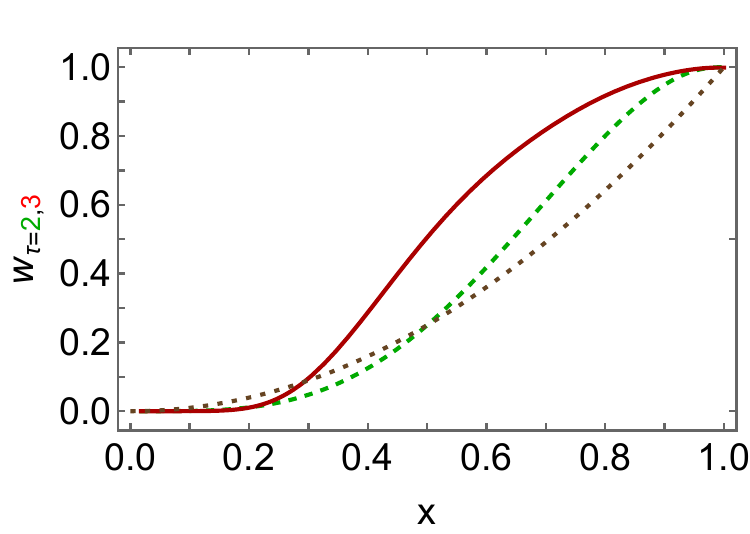}
    \caption{\label{winputN}
    Profile functions, $w_{\tau}(x)$, for the pion ($\tau=2$ -- dashed green) and proton ($\tau=3$ -- solid red) associated with the hadron scale valence quark DFs in Fig.\,\ref{DFinput}.
    Profile associated with Eq.\,\eqref{SCIDF} -- dotted brown curve.
    }
\end{figure}

Using the DFs drawn in Fig.\,\ref{DFinput}, it is straightforward to solve Eq.\,\eqref{dew} numerically and obtain the required profile functions.  The solutions are drawn in Fig.\,\ref{winputN}.  Evidently, they satisfy the constraints in Eq.\,\eqref{wcons}.

\begin{figure}[t]
\vspace*{0ex}

\leftline{\hspace*{0.5em}{{\textsf{A}}}}
\vspace*{-4ex}
\centerline{\includegraphics[width=0.9\columnwidth]{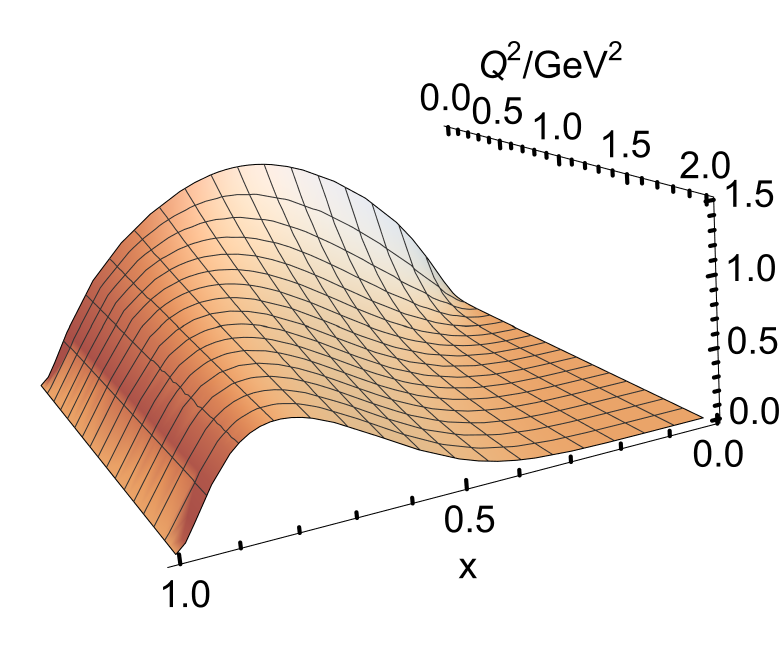}}
\vspace*{0ex}

\leftline{\hspace*{0.5em}{{\textsf{B}}}}
\vspace*{-4ex}
\centerline{\includegraphics[width=0.9\columnwidth]{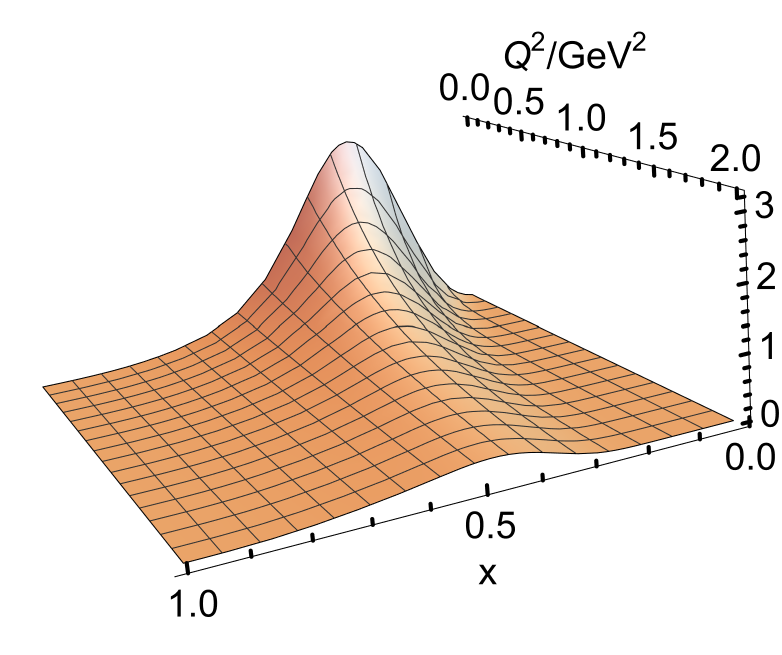}}

\caption{\label{GPDimages}
GPDs (at hadron scale).
Pion ({\sf Panel A}) and proton ({\sf Panel B}), obtained using Eq.\,\eqref{LFHMGPD}, the relevant DFs from Fig.\,\ref{DFinput} and associated profile functions in Fig.\,\ref{winputN}.
At $Q^2=0$, the peaks lie at $x=1/2$, $0.29$, respectively.  They move toward $x=1$ with increasing $Q^2$, showing valence quark dominance of scattering processes at ultraviolet momentum-squared transfers.
}
\end{figure}

At this point, the LFHM pion and proton GPDs are completely specified via Eq.\,\eqref{dewA}.  They are drawn in Fig.\,\ref{GPDimages}: as remarked elsewhere, pion DFs are the most dilated in Nature \cite{Lu:2022cjx}, and this feature is necessarily expressed in GPDs, too.
It is worth reiterating that their forms are entirely fixed by supposing Eq.\,\eqref{eq:EFFint} -- equivalently, Eq.\,\eqref{eq:EFFpoles} -- to be a good approximation to the given hadron's EFF and using CSM predictions for the GPD's forward limit, \emph{viz}.\ the hadron's valence quark DFs at the hadron scale.
The approach taken herein may be understood as an algebraic simplification of the data-driven numerical method for GPD determination employed in Ref.\,\cite{Xu:2023bwv}.
The scale-dependent parton species decompositions of these GPDs can be obtained using the AO evolution scheme, as shown for the pion and kaon in Ref.\,\cite{Raya:2021zrz}.

It should be stressed that owing to the established link with Eq.\,\eqref{eq:EFFint}, the LFHM produces pion and nucleon EFFs that exhibit no sensitivity to the given hadron's valence quark DFs.
The analysis in Ref.\,\cite{Xu:2023bwv} shows that such an outcome is possible with a wide variety of GPD \emph{Ans\"atze}.
However this feature is not a property of QCD because it entails a violation of the connection between valence DF behaviour on $x\simeq 1$ and the large-$Q^2$ scaling-power of hadron form factors \cite{Brodsky:1994kg, Yuan:2003fs, Holt:2010vj, Cui:2021mom, Lu:2022cjx, Lu:2023yna},
\emph{viz}.\ with
${\mathpzc q}_{{\rm v},\tau}(x;\zeta_{\cal H})\sim (1-x)^p$,
$p = 2 \tau -3 +2 \Delta S_z^\tau$, where $\Delta S_z^{2,3}=1/2,0$,
\begin{equation}
F_\tau(Q^2) \sim \left[\frac{1}{Q^2}\right]^{\tau-1}.
\label{DYW}
\end{equation}

The insensitivity to hadron DFs does not persist for higher Mellin moments of a GPD; so, the LFHM prediction for the mass distribution GFF that appears in a hadron's gravitational current is an independent test of the model.
On the other hand, one should not be surprised to find that the sensitivity is weak.


\section{Gravitational form factors}
\label{s3}
The GFF describing the distribution of mass within a hadron may be obtained via \cite{Diehl:2003ny}:
\begin{equation}
A_\tau^q(Q^2) = \int_0^1 dx \, x \, H_{\rm v}(x,0,-Q^2)\,.
\label{AQ2eq}
\end{equation}
In the LFHM context, the GPD is that associated with $\zeta_{\cal H}$.  Additional information, pertaining, \emph{e.g}., to parton species decompositions of GFFs, may be found elsewhere \cite{Raya:2021zrz, Yao:2024ixu}.

\begin{figure}[t]
\vspace*{0ex}

\leftline{\hspace*{0.5em}{{\textsf{A}}}}
\vspace*{-2ex}
\centerline{\includegraphics[width=0.9\columnwidth]{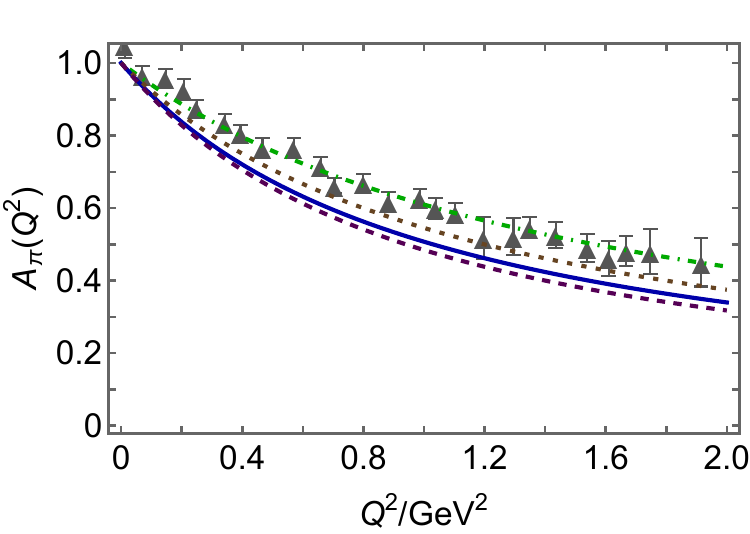}}
\vspace*{0ex}

\leftline{\hspace*{0.5em}{{\textsf{B}}}}
\vspace*{-2ex}
\centerline{\includegraphics[width=0.9\columnwidth]{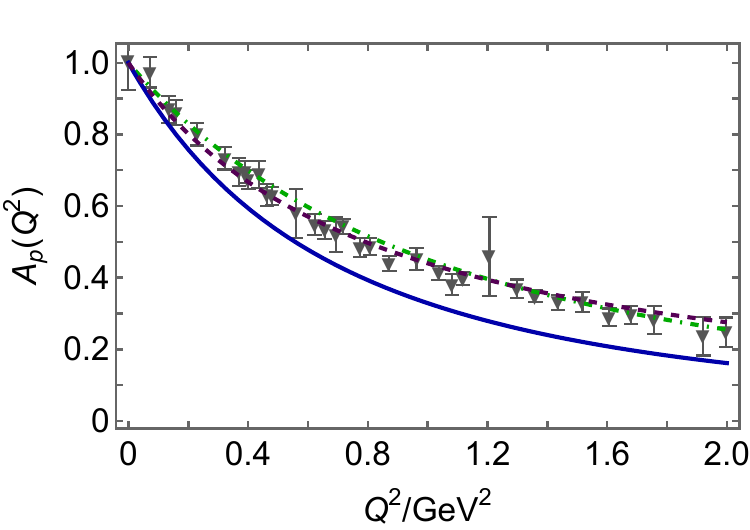}}

\caption{\label{AFF}
{\sf Panel A}.
Pion mass distribution form factor.
Solid blue curve -- LFHM result obtained using Eq.\,\eqref{LFHMGPD} $\tau=2$ solutions in Eq.\,\eqref{AQ2eq};
dashed purple curve -- CSM prediction \cite{Xu:2023izo};
and grey up-triangles -- lQCD results from Ref.\,\cite{Hackett:2023nkr}, obtained using $m_\pi^{\rm lQCD}  \approx 0.17\,$GeV configurations.  (Empirically, $m_\pi = 0.138\,$GeV \cite{ParticleDataGroup:2024cfk}.)
Dotted brown curve -- result obtained using the DF in Eq.\,\eqref{SCIDF}.
%
{\sf Panel B}.
Proton mass distribution form factor.
Solid blue curve -- LFHM obtained using Eq.\,\eqref{LFHMGPD} $\tau=3$ solutions in Eq.\,\eqref{AQ2eq};
dashed purple curve -- CSM prediction \cite{Yao:2024ixu}
and grey down-triangles -- lQCD results from Ref.\,\cite{Hackett:2023rif}, obtained using $m_\pi^{\rm lQCD}  \approx 0.17\,$GeV configurations.
Both panels.
Dot-dashed green curve obtained by replacing $\lambda \to (m_\pi^{\rm lQCD}/m_\pi)^2 \lambda$ in LFHM.
}
\end{figure}

Using the GPDs constructed in Sec.\,\ref{s2} -- depicted in Fig.\,\ref{GPDimages}, one obtains the LFHM results for pion and proton mass distribution GFFs drawn in Fig.\,\ref{AFF}.
In this case, the LFHM pion result is slightly harder than the CSM prediction \cite{Xu:2023izo}, matching the EFF pattern, but this is reversed for the proton, with the LFHM result being softer than the CSM prediction \cite{Yao:2024ixu}.
Relative to the associated electric radii, the pion mass GFF radii are given by:
\begin{align}
                            & \mbox{\rm LFHM} & \mbox{\rm CSM \cite{Xu:2023izo}} \nonumber \\[-1ex]
(r_A^\pi/r_F^\pi)^2 \qquad & 0.58  & 0.54\phantom{[24]\,} \,,
\end{align}
consistent with the bounds in Eq.\,\eqref{radiibound}.


Turning to the proton, a comparison between mass radii and charge radii requires knowledge of the so-called $D$-term in the first case and the Pauli-term in the second:
{\allowdisplaybreaks
\begin{subequations}
\label{protonradii}
\begin{align}
(r_M^p)^2 & = -6 \left. \frac{d}{dQ^2} A(Q^2) \right|_{Q^2=0} - \frac{3}{2 m_p^2} D(0)\,,
\label{protonmass}\\
(r_{\rm ch}^p)^2 & = - 6 \left. \frac{d}{dQ^2} F_{1}^p(Q^2) \right|_{Q^2=0} + \frac{3}{2 m_p^2} F_2^p(0)\,,
\label{protoncharge}
\end{align}
\end{subequations}
where 
$D(0)<0$ is the proton $D$-term
and $F_2^p(0)$ is the proton anomalous magnetic moment.
The Pauli term in Eq.\,\eqref{protoncharge} can be estimated from empirical information: $F_2^p(0) \, 3/[2 m_p^2]=(0.345 {\rm fm})^2$.
Concerning Eq.\,\eqref{protonmass}, the value of $D(0)$ is the ``last unknown global property'' of the proton \cite{Polyakov:2018zvc}.  Using the recent CSM prediction, \emph{viz}.\ $D(0)=-3.11(1)$ \cite{Yao:2024ixu}, one finds
$D(0) \, 3/[2 m_p^2]=-(0.455(1) {\rm fm})^2$.
For the purpose of illustration, we combine these results with the LFHM values of the first terms in Eqs.\,\eqref{protonradii} and obtain:
\begin{align}
 & \mbox{\rm LFHM} & \mbox{\rm CSM \cite{Yao:2024ixu}} \nonumber \\[-1ex]
(r_{\rm mass}^p/r_{\rm ch}^p)^2 \qquad & 0.87  & 0.81(5)\phantom{2\,} \,.
\end{align}
}

Figure\ \ref{AFF} also displays lattice-QCD (lQCD) results, reproduced from Refs.\,\cite{Hackett:2023nkr, Hackett:2023rif}.
Since the CSM predictions are parameter-free, are part of a body of work that unifies pion, kaon, and nucleon electromagnetic and gravitational form factors \cite{Xu:2023izo, Yao:2024drm, Yao:2024ixu}, and agree with data wherever it is available, the displayed comparisons suggest that the lQCD pion results are too hard, possibly in part because the simulation pion mass-squared is 50\% too large.
To check this possibility, we recall that, on a material, relevant domain, pseudoscalar meson charge radii, $r_{0^-}$, satisfy the following relation \cite{Chen:2018rwz}: $r_{0^-}^2 m_{0^-}^2 \approx {\rm constant}$, where $m_{0^-}$ is the meson mass.
One might therefrom argue that in LFHM comparisons with studies that use a heavy pion, the value of $\lambda$ should be increased because it determines the meson radius via Eq.\,\eqref{eq:EFFpoles}.
Implementing $\lambda \to (m_\pi^{\rm lQCD}/m_\pi)^2 \lambda$, one obtains the dot-dashed green curve in Fig.\,\ref{AFF}A, which is a good match for the lQCD result.
The same outcome is achieved for the proton -- see Fig.\,\ref{AFF}B.
Implementing an analogous modification in CSM analyses is not straightforward because hadron charge radii are not so trivially related to the mass of the lightest vector meson.
Consequently, the most that can now reliably be stated is that lQCD results at the physical pion mass are desirable.

It is worth illustrating the sensitivity of LFHM results for $A_\tau$ to the form of the DF.  This is most straightforward for the pion, whose hadron-scale valence quark DF is symmetric around $x=1/2$ -- see Fig.\,\ref{DFinput}.   Consider, therefore, the following DF:
\begin{equation}
\label{SCIDF}
{\mathpzc u}_{2}^{\rm CI}(x;\zeta_{\cal H}) =
1.0202 [x(1-x)]^{0.01}\,.
\end{equation}
Drawn as the dotted brown curve in Fig.\,\ref{DFinput}, it is practically $x$-independent and, thus, very unlike the CSM prediction for ${\mathpzc u}_{\rm v}^\pi(x;\zeta_{\cal H})$: the relative ${\mathpzc L}_1$ difference between these two curves is 38\%.

The profile function determined from Eq.\,\eqref{SCIDF} is drawn as the dotted brown curve in Fig.\,\ref{winputN}.  Unlike the CSM profile function, it is monotic and convex.  In fact, for all practical purposes, one may represent this profile as $w(x) = x^2$.  (This is the solution of Eq.\,\eqref{dew} when ${\mathpzc q}_2(x;Q^2)=1$.)
In that case, the following function is an accurate approximation to the result obtained from Eq.\,\eqref{AQ2eq}:
\begin{equation}
A_{\rm CI}(Q^2) =
2 B(1 , 2+Q^2/2 \lambda)\,,
\end{equation}
which is drawn as the dotted brown curve in Fig.\,\ref{AFF}A.  On the domain displayed, the relative ${\mathpzc L}_1$ difference between $A_{\pi,{\rm CI}}$ is just 6\%.
Evidently, therefore, as anticipated after Eq.\,\eqref{DYW}, in this LFHM approach to building GPDs from EFFs and valence quark DFs, the mass GFF possesses little sensitivity to the input DF because the EFF possesses none.

It is now plain that $A_p(Q^2)$ must be practically independent of the choice made for ${\mathpzc q}_{\tau=3}$; especially, whether one uses ${\mathpzc u}_{\rm v}^p$ or ${\mathpzc d}_{\rm v}^p$ or any weighted average of these functions.

With the charge and mass distribution form factors in hand, one can calculate transverse densities via two-dimensional Fourier transforms \cite{Miller:2010nz}:
\begin{equation}
\label{TDdensity}
\rho_{\cal F}(b_\perp) = \int_0^\infty \frac{dQ}{2\pi} Q J_0(Q b_\perp) {\cal F}(Q^2)\,,
\end{equation}
where $J_0$ is a Bessel function of the first kind, and ${\cal F}=A$ delivers the mass density and ${\cal F}=F$, the charge density.
Owing to the absence of QCD scaling violations in LFHM pion form factors, the analogous three-dimensional Fourier transform is not finite.

\begin{figure}[t]
    \centering
    \includegraphics[width=0.9\columnwidth]{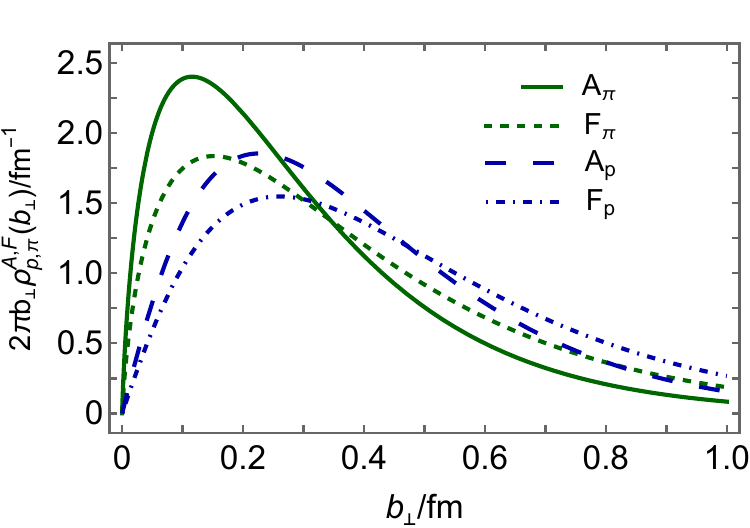}
    \caption{\label{Fdensity}
    Two-dimensional density profiles, calculated using Eq.\,\eqref{TDdensity} for the pion and proton mass and charge distributions.
    }
\end{figure}

The pion and proton profiles are drawn in Fig.\,\ref{Fdensity}.  Evidently, each pion profile is more compressed than the like proton profile -- unsurprisingly, perhaps, given their respective two vs. three valence body content.  Moreover, in each system, the mass profile peaks at a smaller value of $b_\perp$ than the electromagnetic profile and, naturally, given applicable conservation laws, the peak height is greater:
\begin{equation}
\begin{array}{ccccc}
& \multicolumn{2}{c}{b_\perp^{\rm max}/{\rm fm}} & \multicolumn{2}{c}{{\rm height}\times{\rm fm}}   \\
& A & F & A & F\\
\pi & 0.116\  & 0.151\ & 2.40\ & 1.84\ \\
p & 0.226\ & 0.260\ & 1.85\ & 1.55
\end{array}\,.
\end{equation}
For the pion, such outcomes were anticipated in Refs.\,\cite{Kumano:2017lhr, Xu:2023izo}; and they have since been found in continuum Schwinger function studies of proton form factors \cite{Yao:2024uej, Yao:2024ixu}.

It should be borne in mind that since hadron wave functions are independent of the probe used to explore them, then the relative ordering of mass and electric radii is a statement about the probe itself, highlighting that distinct probes perceive different features of the target constituents.
In the present discussion, one notes that a target dressed-quark carries the same charge, irrespective of its momentum.  So, the hadron wave function alone controls the distribution of charge.
On the other hand, the gravitational interaction of a target dressed-quark depends on its momentum – see Eq.\,\eqref{AQ2eq}; after all, the current relates to the energy momentum tensor.
Hence, the hadron's mass GFF depends on interference between quark momentum growth
and wave function momentum suppression.
This pushes support to a larger momentum domain in the hadron, which corresponds to a smaller distance domain.


\section{Summary and Outlook}
\label{s4}
Within the context of a light-front holographic model of hadrons (LFHM), we presented an algebraic method that enables one to construct a fair approximation to a hadron's zero-skewness GPD from its hadron-scale valence-quark distribution function (DF) and electromagnetic form factor (EFF), without reference to scarce relevant data on deeply virtual Compton scattering.  The approach lies within a class that exploits a separable form for the hadron's light-front wave function \cite{Xu:2023bwv}.  Such schemes typically do not preserve the connection between valence DF behaviour on $x\simeq 1$ and the large-$Q^2$ scaling-power of hadron form factors [Eq.\,\eqref{DYW}], although that can be implemented as a constraint on their construction, as was done herein.

Using the hadron GPD built in this way [Fig.\,\ref{GPDimages}], one can immediately calculate the hadron's mass distribution gravitational form factor (GFF), $A_H$ [Fig.\,\ref{AFF}].  Its sensitivity to the hadron's DF is weak; but since that DF is merely the GPDs forward limit, this is possibly a general feature.  The $Q^2$-dependence of $A_H$ is most closely tied to that of the related EFF, $F_H$.  For both the pion and proton, $A_H$ is harder than $F_H$, \emph{i.e}., falls less rapidly with increasing $Q^2$.  Consequently, the related light-front transverse plane mass density profile is more compact than the analogous charge density profile, with each pion near core density being larger than that of its proton partner [Fig.\,\ref{Fdensity}].
These outcomes are not a statement about the target wave function; instead, they are an expression of those qualities of the target constituent to which the different probes are sensitive.

It would be natural to extend this analysis to spin dependent aspects of the proton, relating helicity dependent DFs and spin-flip EFFs to $J(Q^2)$, the proton's spin distribution GFF.  The approach could also be adapted to more complicated bound-states, including pion and nucleon excitations.

%
\medskip

\noindent\textbf{Acknowledgments}.
We are grateful to the authors of Refs.\,\cite{Hackett:2023nkr, Hackett:2023rif} for supplying us with the lattice-QCD results used in Fig.\,\ref{AFF}.
%
Work supported by:
National Natural Science Foundation of China (grant no.\ 12135007);
Spanish Ministry of Science and Innovation (MICINN grant no.\ PID2022-140440NB-C22);
and
Junta de Andaluc{\'{\i}}a (grant no.\ P18-FR-5057).

\medskip
\noindent\textbf{Data Availability Statement}. This manuscript has no associated data or the data will not be deposited. [Authors' comment: All information necessary to reproduce the results described herein is contained in the material presented above.]

\medskip
\noindent\textbf{Declaration of Competing Interest}.
The authors declare that they have no known competing financial interests or personal relationships that could have appeared to influence the work reported in this paper.


\begin{thebibliography}{62}
\providecommand{\natexlab}[1]{#1}
\providecommand{\url}[1]{\texttt{#1}}
\providecommand{\urlprefix}{URL }
\expandafter\ifx\csname urlstyle\endcsname\relax
  \providecommand{\doi}[1]{doi:\discretionary{}{}{}#1}\else
  \providecommand{\doi}[1]{doi:\discretionary{}{}{}\begingroup
  \urlstyle{rm}\url{#1}\endgroup}\fi
\providecommand{\bibinfo}[2]{#2}

\bibitem[{Perdrisat et~al.(2007)Perdrisat, Punjabi, and
  Vanderhaeghen}]{Perdrisat:2006hj}
\bibinfo{author}{C.~F. Perdrisat}, \bibinfo{author}{V.~Punjabi},
  \bibinfo{author}{M.~Vanderhaeghen}, \bibinfo{title}{{Nucleon electromagnetic
  form factors}}, \bibinfo{journal}{Prog. Part. Nucl. Phys.}
  \bibinfo{volume}{59} (\bibinfo{year}{2007}) \bibinfo{pages}{694--764}.

\bibitem[{Eichmann et~al.(2016)Eichmann, Sanchis-Alepuz, Williams, Alkofer, and
  Fischer}]{Eichmann:2016yit}
\bibinfo{author}{G.~Eichmann}, \bibinfo{author}{H.~Sanchis-Alepuz},
  \bibinfo{author}{R.~Williams}, \bibinfo{author}{R.~Alkofer},
  \bibinfo{author}{C.~S. Fischer}, \bibinfo{title}{{Baryons as relativistic
  three-quark bound states}}, \bibinfo{journal}{Prog. Part. Nucl. Phys.}
  \bibinfo{volume}{91} (\bibinfo{year}{2016}) \bibinfo{pages}{1--100}.

\bibitem[{Brodsky et~al.(2020)}]{Brodsky:2020vco}
\bibinfo{author}{S.~J. Brodsky}, et~al., \bibinfo{title}{{Strong QCD from
  Hadron Structure Experiments}}, \bibinfo{journal}{Int. J. Mod. Phys. E}
  \bibinfo{volume}{29}~(\bibinfo{number}{08}) (\bibinfo{year}{2020})
  \bibinfo{pages}{2030006}.

\bibitem[{Roberts et~al.(2021)Roberts, Richards, Horn, and
  Chang}]{Roberts:2021nhw}
\bibinfo{author}{C.~D. Roberts}, \bibinfo{author}{D.~G. Richards},
  \bibinfo{author}{T.~Horn}, \bibinfo{author}{L.~Chang},
  \bibinfo{title}{{Insights into the emergence of mass from studies of pion and
  kaon structure}}, \bibinfo{journal}{Prog. Part. Nucl. Phys.}
  \bibinfo{volume}{120} (\bibinfo{year}{2021}) \bibinfo{pages}{103883}.

\bibitem[{Arrington et~al.(2021)}]{Arrington:2021biu}
\bibinfo{author}{J.~Arrington}, et~al., \bibinfo{title}{{Revealing the
  structure of light pseudoscalar mesons at the electron\textendash{}ion
  collider}}, \bibinfo{journal}{J. Phys. G} \bibinfo{volume}{48}
  (\bibinfo{year}{2021}) \bibinfo{pages}{075106}.

\bibitem[{Chen et~al.(2020)Chen, Guo, Roberts, and Wang}]{Chen:2020ijn}
\bibinfo{author}{X.~Chen}, \bibinfo{author}{F.-K. Guo}, \bibinfo{author}{C.~D.
  Roberts}, \bibinfo{author}{R.~Wang}, \bibinfo{title}{{Selected Science
  Opportunities for the EicC}}, \bibinfo{journal}{Few Body Syst.}
  \bibinfo{volume}{61} (\bibinfo{year}{2020}) \bibinfo{pages}{43}.

\bibitem[{Wang and Chen(2022)}]{Wang:2022xad}
\bibinfo{author}{R.~Wang}, \bibinfo{author}{X.~Chen}, \bibinfo{title}{{The
  Current Status of Electron Ion Collider in China}}, \bibinfo{journal}{Few
  Body Syst.} \bibinfo{volume}{63}~(\bibinfo{number}{2}) (\bibinfo{year}{2022})
  \bibinfo{pages}{48}.

\bibitem[{Carman et~al.(2023)Carman, Gothe, Mokeev, and
  Roberts}]{Carman:2023zke}
\bibinfo{author}{D.~S. Carman}, \bibinfo{author}{R.~W. Gothe},
  \bibinfo{author}{V.~I. Mokeev}, \bibinfo{author}{C.~D. Roberts},
  \bibinfo{title}{{Nucleon Resonance Electroexcitation Amplitudes and Emergent
  Hadron Mass}}, \bibinfo{journal}{Particles}
  \bibinfo{volume}{6}~(\bibinfo{number}{1}) (\bibinfo{year}{2023})
  \bibinfo{pages}{416--439}.

\bibitem[{Ji(1997)}]{Ji:1996ek}
\bibinfo{author}{X.-D. Ji}, \bibinfo{title}{{Gauge-Invariant Decomposition of
  Nucleon Spin}}, \bibinfo{journal}{Phys. Rev. Lett.} \bibinfo{volume}{78}
  (\bibinfo{year}{1997}) \bibinfo{pages}{610--613}.

\bibitem[{Radyushkin(1996)}]{Radyushkin:1996nd}
\bibinfo{author}{A.~Radyushkin}, \bibinfo{title}{{Scaling limit of deeply
  virtual Compton scattering}}, \bibinfo{journal}{Phys. Lett. B}
  \bibinfo{volume}{380} (\bibinfo{year}{1996}) \bibinfo{pages}{417--425}.

\bibitem[{Mueller et~al.(1994)Mueller, Robaschik, Geyer, Dittes, and
  Ho\v{r}ej\v{s}i}]{Mueller:1998fv}
\bibinfo{author}{D.~Mueller}, \bibinfo{author}{D.~Robaschik},
  \bibinfo{author}{B.~Geyer}, \bibinfo{author}{F.~M. Dittes},
  \bibinfo{author}{J.~Ho\v{r}ej\v{s}i}, \bibinfo{title}{{Wave functions,
  evolution equations and evolution kernels from light ray operators of QCD}},
  \bibinfo{journal}{Fortschr. Phys.} \bibinfo{volume}{42}
  (\bibinfo{year}{1994}) \bibinfo{pages}{101}.

\bibitem[{Diehl(2003)}]{Diehl:2003ny}
\bibinfo{author}{M.~Diehl}, \bibinfo{title}{{Generalized parton
  distributions}}, \bibinfo{journal}{Phys. Rept.} \bibinfo{volume}{388}
  (\bibinfo{year}{2003}) \bibinfo{pages}{41--277}.

\bibitem[{Guidal et~al.(2013)Guidal, Moutarde, and
  Vanderhaeghen}]{Guidal:2013rya}
\bibinfo{author}{M.~Guidal}, \bibinfo{author}{H.~Moutarde},
  \bibinfo{author}{M.~Vanderhaeghen}, \bibinfo{title}{{Generalized Parton
  Distributions in the valence region from Deeply Virtual Compton Scattering}},
  \bibinfo{journal}{Rept. Prog. Phys.} \bibinfo{volume}{76}
  (\bibinfo{year}{2013}) \bibinfo{pages}{066202}.

\bibitem[{Mezrag(2023)}]{Mezrag:2023nkp}
\bibinfo{author}{C.~Mezrag}, \bibinfo{title}{{Generalised Parton Distributions
  in Continuum Schwinger Methods: Progresses, Opportunities and Challenges}},
  \bibinfo{journal}{Particles} \bibinfo{volume}{6}~(\bibinfo{number}{1})
  (\bibinfo{year}{2023}) \bibinfo{pages}{262--296}.

\bibitem[{Polyakov and Schweitzer(2018)}]{Polyakov:2018zvc}
\bibinfo{author}{M.~V. Polyakov}, \bibinfo{author}{P.~Schweitzer},
  \bibinfo{title}{{Forces inside hadrons: pressure, surface tension, mechanical
  radius, and all that}}, \bibinfo{journal}{Int. J. Mod. Phys. A}
  \bibinfo{volume}{33}~(\bibinfo{number}{26}) (\bibinfo{year}{2018})
  \bibinfo{pages}{1830025}.

\bibitem[{Azizi and \"Ozdem(2020)}]{Azizi:2019ytx}
\bibinfo{author}{K.~Azizi}, \bibinfo{author}{U.~\"Ozdem},
  \bibinfo{title}{{Nucleon\textquoteright{}s energy\textendash{}momentum tensor
  form factors in light-cone QCD}}, \bibinfo{journal}{Eur. Phys. J. C}
  \bibinfo{volume}{80}~(\bibinfo{number}{2}) (\bibinfo{year}{2020})
  \bibinfo{pages}{104}.

\bibitem[{Neubelt et~al.(2020)Neubelt, Sampino, Hudson, Tezgin, and
  Schweitzer}]{Neubelt:2019sou}
\bibinfo{author}{M.~J. Neubelt}, \bibinfo{author}{A.~Sampino},
  \bibinfo{author}{J.~Hudson}, \bibinfo{author}{K.~Tezgin},
  \bibinfo{author}{P.~Schweitzer}, \bibinfo{title}{{Energy momentum tensor and
  the D-term in the bag model}}, \bibinfo{journal}{Phys. Rev. D}
  \bibinfo{volume}{101}~(\bibinfo{number}{3}) (\bibinfo{year}{2020})
  \bibinfo{pages}{034013}.

\bibitem[{Xing et~al.(2023)Xing, Ding, and Chang}]{Xing:2022mvk}
\bibinfo{author}{Z.~Xing}, \bibinfo{author}{M.~Ding},
  \bibinfo{author}{L.~Chang}, \bibinfo{title}{{Glimpse into the pion
  gravitational form factor}}, \bibinfo{journal}{Phys. Rev. D}
  \bibinfo{volume}{107}~(\bibinfo{number}{3}) (\bibinfo{year}{2023})
  \bibinfo{pages}{L031502}.

\bibitem[{Shi et~al.(2023)Shi, Li, Yin, and Jia}]{Shi:2023oll}
\bibinfo{author}{C.~Shi}, \bibinfo{author}{J.~Li}, \bibinfo{author}{P.-L. Yin},
  \bibinfo{author}{W.~Jia}, \bibinfo{title}{{Unpolarized generalized parton
  distributions of light and heavy vector mesons}}, \bibinfo{journal}{Phys.
  Rev. D} \bibinfo{volume}{107}~(\bibinfo{number}{7}) (\bibinfo{year}{2023})
  \bibinfo{pages}{074009}.

\bibitem[{Li and Vary(2024)}]{Li:2023izn}
\bibinfo{author}{Y.~Li}, \bibinfo{author}{J.~P. Vary}, \bibinfo{title}{{Stress
  inside the pion in holographic light-front QCD}}, \bibinfo{journal}{Phys.
  Rev. D} \bibinfo{volume}{109}~(\bibinfo{number}{5}) (\bibinfo{year}{2024})
  \bibinfo{pages}{L051501}.

\bibitem[{Nair et~al.(2024)Nair, Mondal, Xu, Zhao, Mukherjee, and
  Vary}]{Nair:2024fit}
\bibinfo{author}{S.~Nair}, \bibinfo{author}{C.~Mondal},
  \bibinfo{author}{S.~Xu}, \bibinfo{author}{X.~Zhao},
  \bibinfo{author}{A.~Mukherjee}, \bibinfo{author}{J.~P. Vary},
  \bibinfo{title}{{Gravitational form factors and mechanical properties of
  quarks in protons: A basis light-front quantization approach}},
  \bibinfo{journal}{Phys. Rev. D} \bibinfo{volume}{110}~(\bibinfo{number}{5})
  (\bibinfo{year}{2024}) \bibinfo{pages}{056027}.

\bibitem[{Liu et~al.(2024)Liu, Shuryak, and Zahed}]{Liu:2024vkj}
\bibinfo{author}{W.-Y. Liu}, \bibinfo{author}{E.~Shuryak},
  \bibinfo{author}{I.~Zahed}, \bibinfo{title}{{Pion gravitational form factors
  in the QCD instanton vacuum. II}}, \bibinfo{journal}{Phys. Rev. D}
  \bibinfo{volume}{110}~(\bibinfo{number}{5}) (\bibinfo{year}{2024})
  \bibinfo{pages}{054022}.

\bibitem[{Xu et~al.(2024)Xu, Ding, Raya, Roberts, Rodr\'\i{}guez-Quintero, and
  Schmidt}]{Xu:2023izo}
\bibinfo{author}{Y.-Z. Xu}, \bibinfo{author}{M.~Ding},
  \bibinfo{author}{K.~Raya}, \bibinfo{author}{C.~D. Roberts},
  \bibinfo{author}{J.~Rodr\'\i{}guez-Quintero}, \bibinfo{author}{S.~M.
  Schmidt}, \bibinfo{title}{{Pion and kaon electromagnetic and gravitational
  form factors}}, \bibinfo{journal}{Eur. Phys. J. C}
  \bibinfo{volume}{84}~(\bibinfo{number}{2}) (\bibinfo{year}{2024})
  \bibinfo{pages}{191}.

\bibitem[{Yao et~al.(2024{\natexlab{a}})Yao, Xu, Binosi, Cui, Ding, Raya,
  Roberts, Rodr\'\i{}guez-Quintero, and Schmidt}]{Yao:2024ixu}
\bibinfo{author}{Z.~Q. Yao}, \bibinfo{author}{Y.~Z. Xu},
  \bibinfo{author}{D.~Binosi}, \bibinfo{author}{Z.~F. Cui},
  \bibinfo{author}{M.~Ding}, \bibinfo{author}{K.~Raya}, \bibinfo{author}{C.~D.
  Roberts}, \bibinfo{author}{J.~Rodr\'\i{}guez-Quintero},
  \bibinfo{author}{S.~M. Schmidt}, \bibinfo{title}{{Nucleon Gravitational Form
  Factors -- arXiv:2409.15547 [hep-ph]}} .

\bibitem[{Hackett et~al.(2024)Hackett, Pefkou, and Shanahan}]{Hackett:2023rif}
\bibinfo{author}{D.~C. Hackett}, \bibinfo{author}{D.~A. Pefkou},
  \bibinfo{author}{P.~E. Shanahan}, \bibinfo{title}{{Gravitational Form Factors
  of the Proton from Lattice QCD}}, \bibinfo{journal}{Phys. Rev. Lett.}
  \bibinfo{volume}{132}~(\bibinfo{number}{25}) (\bibinfo{year}{2024})
  \bibinfo{pages}{251904}.

\bibitem[{Hackett et~al.(2023)Hackett, Oare, Pefkou, and
  Shanahan}]{Hackett:2023nkr}
\bibinfo{author}{D.~C. Hackett}, \bibinfo{author}{P.~R. Oare},
  \bibinfo{author}{D.~A. Pefkou}, \bibinfo{author}{P.~E. Shanahan},
  \bibinfo{title}{{Gravitational form factors of the pion from lattice QCD}},
  \bibinfo{journal}{Phys. Rev. D} \bibinfo{volume}{108}~(\bibinfo{number}{11})
  (\bibinfo{year}{2023}) \bibinfo{pages}{114504}.

\bibitem[{Binosi(2022)}]{Binosi:2022djx}
\bibinfo{author}{D.~Binosi}, \bibinfo{title}{{Emergent Hadron Mass in Strong
  Dynamics}}, \bibinfo{journal}{Few Body Syst.}
  \bibinfo{volume}{63}~(\bibinfo{number}{2}) (\bibinfo{year}{2022})
  \bibinfo{pages}{42}.

\bibitem[{Ding et~al.(2023)Ding, Roberts, and Schmidt}]{Ding:2022ows}
\bibinfo{author}{M.~Ding}, \bibinfo{author}{C.~D. Roberts},
  \bibinfo{author}{S.~M. Schmidt}, \bibinfo{title}{{Emergence of Hadron Mass
  and Structure}}, \bibinfo{journal}{Particles}
  \bibinfo{volume}{6}~(\bibinfo{number}{1}) (\bibinfo{year}{2023})
  \bibinfo{pages}{57--120}.

\bibitem[{Ferreira and Papavassiliou(2023)}]{Ferreira:2023fva}
\bibinfo{author}{M.~N. Ferreira}, \bibinfo{author}{J.~Papavassiliou},
  \bibinfo{title}{{Gauge Sector Dynamics in QCD}}, \bibinfo{journal}{Particles}
  \bibinfo{volume}{6}~(\bibinfo{number}{1}) (\bibinfo{year}{2023})
  \bibinfo{pages}{312--363}.

\bibitem[{Raya et~al.(2024)Raya, Bashir, Binosi, Roberts, and
  Rodr\'\i{}guez-Quintero}]{Raya:2024ejx}
\bibinfo{author}{K.~Raya}, \bibinfo{author}{A.~Bashir},
  \bibinfo{author}{D.~Binosi}, \bibinfo{author}{C.~D. Roberts},
  \bibinfo{author}{J.~Rodr\'\i{}guez-Quintero}, \bibinfo{title}{{Pseudoscalar
  Mesons and Emergent Mass}}, \bibinfo{journal}{Few Body Syst.}
  \bibinfo{volume}{65}~(\bibinfo{number}{2}) (\bibinfo{year}{2024})
  \bibinfo{pages}{60}.

\bibitem[{Xu et~al.(2023)Xu, Raya, Cui, Roberts, and
  Rodr\'\i{}guez-Quintero}]{Xu:2023bwv}
\bibinfo{author}{Y.-Z. Xu}, \bibinfo{author}{K.~Raya}, \bibinfo{author}{Z.-F.
  Cui}, \bibinfo{author}{C.~D. Roberts},
  \bibinfo{author}{J.~Rodr\'\i{}guez-Quintero}, \bibinfo{title}{{Empirical
  Determination of the Pion Mass Distribution}}, \bibinfo{journal}{Chin. Phys.
  Lett. \emph{Express}} \bibinfo{volume}{40}~(\bibinfo{number}{4})
  (\bibinfo{year}{2023}) \bibinfo{pages}{041201}.

\bibitem[{Conway et~al.(1989)}]{Conway:1989fs}
\bibinfo{author}{J.~S. Conway}, et~al., \bibinfo{title}{{Experimental study of
  muon pairs produced by 252-GeV pions on tungsten}}, \bibinfo{journal}{Phys.
  Rev. D} \bibinfo{volume}{39} (\bibinfo{year}{1989}) \bibinfo{pages}{92--122}.

\bibitem[{Horn et~al.(2008)}]{Horn:2007ug}
\bibinfo{author}{T.~Horn}, et~al., \bibinfo{title}{{Scaling study of the pion
  electroproduction cross sections and the pion form factor}},
  \bibinfo{journal}{Phys. Rev. C} \bibinfo{volume}{78} (\bibinfo{year}{2008})
  \bibinfo{pages}{058201}.

\bibitem[{Huber et~al.(2008)}]{Huber:2008id}
\bibinfo{author}{G.~Huber}, et~al., \bibinfo{title}{{Charged pion form-factor
  between {$Q^2 = 0.60\,$GeV$^2$} and {$2.45\,$GeV$^2$}. II. Determination of,
  and results for, the pion form-factor}}, \bibinfo{journal}{Phys. Rev. C}
  \bibinfo{volume}{78} (\bibinfo{year}{2008}) \bibinfo{pages}{045203}.

\bibitem[{Chavez et~al.(2022)Chavez, Bertone, De~Soto~Borrero, Defurne, Mezrag,
  Moutarde, Rodr\'\i{}guez-Quintero, and Segovia}]{Chavez:2021llq}
\bibinfo{author}{J.~M.~M. Chavez}, \bibinfo{author}{V.~Bertone},
  \bibinfo{author}{F.~De~Soto~Borrero}, \bibinfo{author}{M.~Defurne},
  \bibinfo{author}{C.~Mezrag}, \bibinfo{author}{H.~Moutarde},
  \bibinfo{author}{J.~Rodr\'\i{}guez-Quintero}, \bibinfo{author}{J.~Segovia},
  \bibinfo{title}{{Pion generalized parton distributions: A path toward
  phenomenology}}, \bibinfo{journal}{Phys. Rev. D}
  \bibinfo{volume}{105}~(\bibinfo{number}{9}) (\bibinfo{year}{2022})
  \bibinfo{pages}{094012}.

\bibitem[{Dall'Olio et~al.(2024)Dall'Olio, De~Soto, Mezrag, Morgado~Ch\'avez,
  Moutarde, Rodr\'\i{}guez-Quintero, Sznajder, and Segovia}]{DallOlio:2024vjv}
\bibinfo{author}{P.~Dall'Olio}, \bibinfo{author}{F.~De~Soto},
  \bibinfo{author}{C.~Mezrag}, \bibinfo{author}{J.~M. Morgado~Ch\'avez},
  \bibinfo{author}{H.~Moutarde}, \bibinfo{author}{J.~Rodr\'\i{}guez-Quintero},
  \bibinfo{author}{P.~Sznajder}, \bibinfo{author}{J.~Segovia},
  \bibinfo{title}{{Unraveling generalized parton distributions through Lorentz
  symmetry and partial DGLAP knowledge}}, \bibinfo{journal}{Phys. Rev. D}
  \bibinfo{volume}{109}~(\bibinfo{number}{9}) (\bibinfo{year}{2024})
  \bibinfo{pages}{096013}.

\bibitem[{Sufian et~al.(2017)Sufian, de~T\'eramond, Brodsky, Deur, and
  Dosch}]{Sufian:2016hwn}
\bibinfo{author}{R.~S. Sufian}, \bibinfo{author}{G.~F. de~T\'eramond},
  \bibinfo{author}{S.~J. Brodsky}, \bibinfo{author}{A.~Deur},
  \bibinfo{author}{H.~G. Dosch}, \bibinfo{title}{{Analysis of nucleon
  electromagnetic form factors from light-front holographic QCD : The spacelike
  region}}, \bibinfo{journal}{Phys. Rev. D}
  \bibinfo{volume}{95}~(\bibinfo{number}{1}) (\bibinfo{year}{2017})
  \bibinfo{pages}{014011}.

\bibitem[{de~Teramond et~al.(2018)de~Teramond, Liu, Sufian, Dosch, Brodsky, and
  Deur}]{deTeramond:2018ecg}
\bibinfo{author}{G.~F. de~Teramond}, \bibinfo{author}{T.~Liu},
  \bibinfo{author}{R.~S. Sufian}, \bibinfo{author}{H.~G. Dosch},
  \bibinfo{author}{S.~J. Brodsky}, \bibinfo{author}{A.~Deur},
  \bibinfo{title}{{Universality of Generalized Parton Distributions in
  Light-Front Holographic QCD}}, \bibinfo{journal}{Phys. Rev. Lett.}
  \bibinfo{volume}{120} (\bibinfo{year}{2018}) \bibinfo{pages}{182001}.

\bibitem[{Chang et~al.(2020)Chang, Raya, and Wang}]{Chang:2020kjj}
\bibinfo{author}{L.~Chang}, \bibinfo{author}{K.~Raya},
  \bibinfo{author}{X.~Wang}, \bibinfo{title}{{Pion Parton Distribution Function
  in Light-Front Holographic QCD}}, \bibinfo{journal}{Chin. Phys. C}
  \bibinfo{volume}{44}~(\bibinfo{number}{11}) (\bibinfo{year}{2020})
  \bibinfo{pages}{114105}.

\bibitem[{Navas et~al.(2024)}]{ParticleDataGroup:2024cfk}
\bibinfo{author}{S.~Navas}, et~al., \bibinfo{title}{{Review of particle
  physics}}, \bibinfo{journal}{Phys. Rev. D}
  \bibinfo{volume}{110}~(\bibinfo{number}{3}) (\bibinfo{year}{2024})
  \bibinfo{pages}{030001}.

\bibitem[{Lepage and Brodsky(1980)}]{Lepage:1980fj}
\bibinfo{author}{G.~P. Lepage}, \bibinfo{author}{S.~J. Brodsky},
  \bibinfo{title}{{Exclusive Processes in Perturbative Quantum
  Chromodynamics}}, \bibinfo{journal}{Phys. Rev. D} \bibinfo{volume}{22}
  (\bibinfo{year}{1980}) \bibinfo{pages}{2157--2198}.

\bibitem[{Yao et~al.(2024{\natexlab{b}})Yao, Binosi, and Roberts}]{Yao:2024drm}
\bibinfo{author}{Z.-Q. Yao}, \bibinfo{author}{D.~Binosi},
  \bibinfo{author}{C.~D. Roberts}, \bibinfo{title}{{Onset of scaling violation
  in pion and kaon elastic electromagnetic form factors}},
  \bibinfo{journal}{Phys. Lett. B} \bibinfo{volume}{855}
  (\bibinfo{year}{2024}{\natexlab{b}}) \bibinfo{pages}{138823}.

\bibitem[{Yao et~al.(2024{\natexlab{c}})Yao, Binosi, Cui, and
  Roberts}]{Yao:2024uej}
\bibinfo{author}{Z.-Q. Yao}, \bibinfo{author}{D.~Binosi},
  \bibinfo{author}{Z.-F. Cui}, \bibinfo{author}{C.~D. Roberts},
  \bibinfo{title}{{Nucleon charge and magnetisation distributions: flavour
  separation and zeroes -- arXiv:2403.08088 [hep-ph]}} .

\bibitem[{Kelly(2004)}]{Kelly:2004hm}
\bibinfo{author}{J.~J. Kelly}, \bibinfo{title}{{Simple parametrization of
  nucleon form factors}}, \bibinfo{journal}{Phys. Rev. C} \bibinfo{volume}{70}
  (\bibinfo{year}{2004}) \bibinfo{pages}{068202}.

\bibitem[{Raya et~al.(2022)Raya, Cui, Chang, Morgado, Roberts, and
  Rodr{\'{\i}}guez-Quintero}]{Raya:2021zrz}
\bibinfo{author}{K.~Raya}, \bibinfo{author}{Z.-F. Cui},
  \bibinfo{author}{L.~Chang}, \bibinfo{author}{J.-M. Morgado},
  \bibinfo{author}{C.~D. Roberts},
  \bibinfo{author}{J.~Rodr{\'{\i}}guez-Quintero}, \bibinfo{title}{{Revealing
  pion and kaon structure via generalised parton distributions}},
  \bibinfo{journal}{Chin. Phys. C} \bibinfo{volume}{46}~(\bibinfo{number}{26})
  (\bibinfo{year}{2022}) \bibinfo{pages}{013105}.

\bibitem[{Xu et~al.(2018)Xu, Chang, Roberts, and Zong}]{Xu:2018eii}
\bibinfo{author}{S.-S. Xu}, \bibinfo{author}{L.~Chang}, \bibinfo{author}{C.~D.
  Roberts}, \bibinfo{author}{H.-S. Zong}, \bibinfo{title}{{Pion and kaon
  valence-quark parton quasidistributions}}, \bibinfo{journal}{Phys. Rev. D}
  \bibinfo{volume}{97} (\bibinfo{year}{2018}) \bibinfo{pages}{094014}.

\bibitem[{Yin et~al.(2023)Yin, Xu, Cui, Roberts, and
  Rodr\'\i{}guez-Quintero}]{Yin:2023dbw}
\bibinfo{author}{P.-L. Yin}, \bibinfo{author}{Y.-Z. Xu}, \bibinfo{author}{Z.-F.
  Cui}, \bibinfo{author}{C.~D. Roberts},
  \bibinfo{author}{J.~Rodr\'\i{}guez-Quintero}, \bibinfo{title}{{All-Orders
  Evolution of Parton Distributions: Principle, Practice, and Predictions}},
  \bibinfo{journal}{Chin. Phys. Lett. \emph{Express}}
  \bibinfo{volume}{40}~(\bibinfo{number}{9}) (\bibinfo{year}{2023})
  \bibinfo{pages}{091201}.

\bibitem[{Cui et~al.(2020)Cui, Ding, Gao, Raya, Binosi, Chang, Roberts,
  Rodr\'{\i}guez-Quintero, and Schmidt}]{Cui:2020tdf}
\bibinfo{author}{Z.-F. Cui}, \bibinfo{author}{M.~Ding},
  \bibinfo{author}{F.~Gao}, \bibinfo{author}{K.~Raya},
  \bibinfo{author}{D.~Binosi}, \bibinfo{author}{L.~Chang},
  \bibinfo{author}{C.~D. Roberts},
  \bibinfo{author}{J.~Rodr\'{\i}guez-Quintero}, \bibinfo{author}{S.~M.
  Schmidt}, \bibinfo{title}{{Kaon and pion parton distributions}},
  \bibinfo{journal}{Eur. Phys. J. C} \bibinfo{volume}{80}
  (\bibinfo{year}{2020}) \bibinfo{pages}{1064}.

\bibitem[{Lu et~al.(2022)Lu, Chang, Raya, Roberts, and
  Rodr\'\i{}guez-Quintero}]{Lu:2022cjx}
\bibinfo{author}{Y.~Lu}, \bibinfo{author}{L.~Chang}, \bibinfo{author}{K.~Raya},
  \bibinfo{author}{C.~D. Roberts},
  \bibinfo{author}{J.~Rodr\'\i{}guez-Quintero}, \bibinfo{title}{{Proton and
  pion distribution functions in counterpoint}}, \bibinfo{journal}{Phys. Lett.
  B} \bibinfo{volume}{830} (\bibinfo{year}{2022}) \bibinfo{pages}{137130}.

\bibitem[{Cheng et~al.(2023)Cheng, Yu, Xing, Chen, Cui, and
  Roberts}]{Cheng:2023kmt}
\bibinfo{author}{P.~Cheng}, \bibinfo{author}{Y.~Yu}, \bibinfo{author}{H.-Y.
  Xing}, \bibinfo{author}{C.~Chen}, \bibinfo{author}{Z.-F. Cui},
  \bibinfo{author}{C.~D. Roberts}, \bibinfo{title}{{Perspective on polarised
  parton distribution functions and proton spin}}, \bibinfo{journal}{Phys.
  Lett. B} \bibinfo{volume}{844} (\bibinfo{year}{2023})
  \bibinfo{pages}{138074}.

\bibitem[{Yu et~al.(2024)Yu, Cheng, Xing, Gao, and Roberts}]{Yu:2024qsd}
\bibinfo{author}{Y.~Yu}, \bibinfo{author}{P.~Cheng}, \bibinfo{author}{H.-Y.
  Xing}, \bibinfo{author}{F.~Gao}, \bibinfo{author}{C.~D. Roberts},
  \bibinfo{title}{{Contact interaction study of proton parton distributions}},
  \bibinfo{journal}{Eur. Phys. J. C} \bibinfo{volume}{84}~(\bibinfo{number}{7})
  (\bibinfo{year}{2024}) \bibinfo{pages}{739}.

\bibitem[{Xing et~al.(2024)Xing, Yao, Li, Binosi, Cui, and
  Roberts}]{Xing:2023pms}
\bibinfo{author}{H.~Y. Xing}, \bibinfo{author}{Z.~Q. Yao},
  \bibinfo{author}{B.~L. Li}, \bibinfo{author}{D.~Binosi},
  \bibinfo{author}{Z.~F. Cui}, \bibinfo{author}{C.~D. Roberts},
  \bibinfo{title}{{Developing predictions for pion fragmentation functions}},
  \bibinfo{journal}{Eur. Phys. J. C} \bibinfo{volume}{84}~(\bibinfo{number}{1})
  (\bibinfo{year}{2024}) \bibinfo{pages}{82}.

\bibitem[{Abrams et~al.(2022)}]{Abrams:2021xum}
\bibinfo{author}{D.~Abrams}, et~al., \bibinfo{title}{{Measurement of the
  Nucleon $F^n_2/F^p_2$ Structure Function Ratio by the Jefferson Lab MARATHON
  Tritium/Helium-3 Deep Inelastic Scattering Experiment}},
  \bibinfo{journal}{Phys. Rev. Lett.}
  \bibinfo{volume}{128}~(\bibinfo{number}{13}) (\bibinfo{year}{2022})
  \bibinfo{pages}{132003}.

\bibitem[{Cui et~al.(2022{\natexlab{a}})Cui, Gao, Binosi, Chang, Roberts, and
  Schmidt}]{Cui:2021gzg}
\bibinfo{author}{Z.-F. Cui}, \bibinfo{author}{F.~Gao},
  \bibinfo{author}{D.~Binosi}, \bibinfo{author}{L.~Chang},
  \bibinfo{author}{C.~D. Roberts}, \bibinfo{author}{S.~M. Schmidt},
  \bibinfo{title}{{Valence quark ratio in the proton}}, \bibinfo{journal}{Chin.
  Phys. Lett. \emph{Express}} \bibinfo{volume}{39}~(\bibinfo{number}{04})
  (\bibinfo{year}{2022}{\natexlab{a}}) \bibinfo{pages}{041401}.

\bibitem[{Brodsky et~al.(1995)Brodsky, Burkardt, and Schmidt}]{Brodsky:1994kg}
\bibinfo{author}{S.~J. Brodsky}, \bibinfo{author}{M.~Burkardt},
  \bibinfo{author}{I.~Schmidt}, \bibinfo{title}{{Perturbative QCD constraints
  on the shape of polarized quark and gluon distributions}},
  \bibinfo{journal}{Nucl. Phys. B} \bibinfo{volume}{441} (\bibinfo{year}{1995})
  \bibinfo{pages}{197--214}.

\bibitem[{Yuan(2004)}]{Yuan:2003fs}
\bibinfo{author}{F.~Yuan}, \bibinfo{title}{{Generalized parton distributions at
  $x \to 1$}}, \bibinfo{journal}{Phys. Rev. D} \bibinfo{volume}{69}
  (\bibinfo{year}{2004}) \bibinfo{pages}{051501}.

\bibitem[{Holt and Roberts(2010)}]{Holt:2010vj}
\bibinfo{author}{R.~J. Holt}, \bibinfo{author}{C.~D. Roberts},
  \bibinfo{title}{{Distribution Functions of the Nucleon and Pion in the
  Valence Region}}, \bibinfo{journal}{Rev. Mod. Phys.} \bibinfo{volume}{82}
  (\bibinfo{year}{2010}) \bibinfo{pages}{2991--3044}.

\bibitem[{Cui et~al.(2022{\natexlab{b}})Cui, Ding, Morgado, Raya, Binosi,
  Chang, Papavassiliou, Roberts, Rodr\'\i{}guez-Quintero, and
  Schmidt}]{Cui:2021mom}
\bibinfo{author}{Z.~F. Cui}, \bibinfo{author}{M.~Ding}, \bibinfo{author}{J.~M.
  Morgado}, \bibinfo{author}{K.~Raya}, \bibinfo{author}{D.~Binosi},
  \bibinfo{author}{L.~Chang}, \bibinfo{author}{J.~Papavassiliou},
  \bibinfo{author}{C.~D. Roberts},
  \bibinfo{author}{J.~Rodr\'\i{}guez-Quintero}, \bibinfo{author}{S.~M.
  Schmidt}, \bibinfo{title}{{Concerning pion parton distributions}},
  \bibinfo{journal}{Eur. Phys. J. A} \bibinfo{volume}{58}~(\bibinfo{number}{1})
  (\bibinfo{year}{2022}{\natexlab{b}}) \bibinfo{pages}{10}.

\bibitem[{Lu et~al.(2024)Lu, Xu, Raya, Roberts, and
  Rodr\'\i{}guez-Quintero}]{Lu:2023yna}
\bibinfo{author}{Y.~Lu}, \bibinfo{author}{Y.-Z. Xu}, \bibinfo{author}{K.~Raya},
  \bibinfo{author}{C.~D. Roberts},
  \bibinfo{author}{J.~Rodr\'\i{}guez-Quintero}, \bibinfo{title}{{Pion
  distribution functions from low-order Mellin moments}},
  \bibinfo{journal}{Phys. Lett. B} \bibinfo{volume}{850} (\bibinfo{year}{2024})
  \bibinfo{pages}{138534}.

\bibitem[{Chen et~al.(2018)Chen, Ding, Chang, and Roberts}]{Chen:2018rwz}
\bibinfo{author}{M.~Chen}, \bibinfo{author}{M.~Ding},
  \bibinfo{author}{L.~Chang}, \bibinfo{author}{C.~D. Roberts},
  \bibinfo{title}{{Mass-dependence of pseudoscalar meson elastic form
  factors}}, \bibinfo{journal}{Phys. Rev. D} \bibinfo{volume}{98}
  (\bibinfo{year}{2018}) \bibinfo{pages}{091505(R)}.

\bibitem[{Miller(2010)}]{Miller:2010nz}
\bibinfo{author}{G.~A. Miller}, \bibinfo{title}{{Transverse Charge Densities}},
  \bibinfo{journal}{Ann. Rev. Nucl. Part. Sci.} \bibinfo{volume}{60}
  (\bibinfo{year}{2010}) \bibinfo{pages}{1--25}.

\bibitem[{Kumano et~al.(2018)Kumano, Song, and Teryaev}]{Kumano:2017lhr}
\bibinfo{author}{S.~Kumano}, \bibinfo{author}{Q.-T. Song},
  \bibinfo{author}{O.~V. Teryaev}, \bibinfo{title}{{Hadron tomography by
  generalized distribution amplitudes in pion-pair production process $\gamma^*
  \gamma \rightarrow \pi^0 \pi^0 $ and gravitational form factors for pion}},
  \bibinfo{journal}{Phys. Rev. D} \bibinfo{volume}{97} (\bibinfo{year}{2018})
  \bibinfo{pages}{014020}.

\end{thebibliography}

\end{document}